\documentclass[twocolumn]{aastex62}

\newcommand{\mj}{$\mathrm{M_{Jup}}$}
\newcommand{\rj}{$\mathrm{R_{Jup}}$}

\graphicspath{{./}{figures/}}

\received{December 15, 2025}
\revised{January 12, 2026}
\accepted{\today}
\submitjournal{ApJ}
\usepackage{comment}
\usepackage{gensymb}
\usepackage{lineno}

\usepackage{placeins}
\shorttitle{Direct detection of WISPIT 2c}
\shortauthors{Lawlor et al.}

\begin{document}

\title{Direct spectroscopic confirmation of the young embedded proto-planet WISPIT 2c}

\correspondingauthor{Chloe Lawlor}
\email{chloelawlor766@gmail.com}









\author[0009-0001-0368-1062]{Chloe Lawlor}
\affiliation{School of Natural Sciences, Center for Astronomy,
University of Galway, Galway, H91 CF50, Ireland}
\affiliation{Ryan Institute,
University of Galway, Galway, H91 TK33, Ireland}

\author[0009-0002-6729-646X]{Richelle F. van Capelleveen}
\affiliation{Leiden Observatory, Leiden University,
Postbus 9513, 2300 RA Leiden, The Netherlands}

\author[0000-0002-6777-6386]{Guillaume Bourdarot}
\affiliation{Max Planck Institute for Extraterrestrial Physics,
Gießenbachstraße 1, 85748 Garching, Germany}

\author[0000-0002-4438-1971]{Christian Ginski}
\affiliation{School of Natural Sciences, Center for Astronomy,
University of Galway, Galway, H91 CF50, Ireland}
\affiliation{Ryan Institute,
University of Galway, Galway, H91 TK33, Ireland}

\author[0000-0002-7064-8270]{Matthew A. Kenworthy}
\affiliation{Leiden Observatory, Leiden University,
Postbus 9513, 2300 RA Leiden, The Netherlands}

\author[0000-0002-5823-3072]{Tomas Stolker}
\affiliation{Leiden Observatory, Leiden University,
Postbus 9513, 2300 RA Leiden, The Netherlands}

\author[0000-0002-2167-8246]{Laird Close}
\affiliation{Center for Astronomical Adaptive Optics,
Department of Astronomy, University of Arizona,
933 N. Cherry Avenue, Tucson, AZ 85718, USA}

\author[0000-0003-1401-9952]{Alexander J.~Bohn}
\affiliation{Leiden Observatory, Leiden University, Postbus 9513, 2300 RA Leiden, The Netherlands}

\author{Frank Eisenhauer}
\affiliation{Max Planck Institute for Extraterrestrial Physics,
Gießenbachstraße 1, 85748 Garching, Germany}
\affiliation{Department of Physics,
Technical University of Munich, 85748 Garching, Germany}

\author{Paulo Garcia}
\affiliation{Faculdade de Engenharia, Universidade do Porto,
rua Dr. Roberto Frias, 4200-465 Porto, Portugal}
\affiliation{CENTRA -- Centro de Astrofísica e Gravitação,
IST, Universidade de Lisboa, 1049-001 Lisboa, Portugal}

\author{Sebastian F. Hönig}
\affiliation{School of Physics and Astronomy,
University of Southampton, Southampton, SO17 1BJ, United Kingdom}

\author{Jens Kammerer}
\affiliation{European Southern Observatory, Karl-Schwarzschild-Straße 2,
85748 Garching, Germany}

\author[0000-0003-0514-1147]{Laura Kreidberg}
\affiliation{Max Planck Institute for Astronomy,
Königstuhl 17, 69117 Heidelberg, Germany}

\author[0000-0002-6948-0263]{Sylvestre Lacour}
\affiliation{LIRA, Observatoire de Paris, Université PSL, CNRS,
Sorbonne Université, Université de Paris,
5 place Jules Janssen, 92195 Meudon, France}

\author[0000-0002-0493-4674]{Jean-Baptiste Le Bouquin}
\affiliation{Univ. Grenoble Alpes, CNRS, IPAG,
38000 Grenoble, France}

\author[0000-0003-2008-1488]{Eric Mamajek}
\affiliation{Jet Propulsion Laboratory,
California Institute of Technology,
4800 Oak Grove Drive, Pasadena, CA 91109, USA}

\author{Mathias Nowak}
\affiliation{LIRA, Observatoire de Paris, Université PSL, CNRS,
Sorbonne Université, Université de Paris,
5 place Jules Janssen, 92195 Meudon, France}

\author[0000-0003-0655-0452]{Thibaut Paumard}
\affiliation{LIRA, Observatoire de Paris, Université PSL, CNRS, Sorbonne Université, Université de Paris, 5 place Jules Janssen, 92195 Meudon, France}

\author[0000-0002-0671-9302]{Christian Straubmeier}
\affiliation{1st Institute of Physics,
University of Cologne,
Zülpicher Straße 77, 50937 Cologne, Germany}

\author[0000-0003-2458-9756]{Nienke van der Marel}
\affiliation{Leiden Observatory, Leiden University, Postbus 9513, 2300 RA Leiden, The Netherlands}

\author{the exoGRAVITY Collaboration}



\begin{abstract}

WISPIT 2 is a nearby young star with a multi-ringed disk which was recently confirmed to host a $\sim$ 4.9 \mj{} gas giant planet embedded in a large (60 au) gap at a radial separation of 57 au from the host star.
We confirm and characterise a second, close-in planet in the WISPIT 2 system using a combination of new VLT/SPHERE $H$-band dual-polarisation imaging and VLTI/GRAVITY $K$-band interferometric observations of the WISPIT 2 system.
The GRAVITY detection is consistent with a point-like source while its extracted $K$-band spectrum shows CO band-head absorption at 2.3 $\text{\textmu m}$ and a continuum shape consistent with a young giant planet.
From the GRAVITY data we extract a medium resolution $K$-band spectrum of the companion and fit atmospheric model grids using the \texttt{species} tool with nested sampling to constrain its effective temperature, radius, and luminosity.
We infer T$_\mathrm{eff}$ of 1500-2600 K, a radius of 0.91-2.2 \rj{}, and a luminosity of (-3.47)-(-3.63).
Comparison with evolutionary tracks implies a mass range of 8-12 \mj{}, approximately twice as massive as the previously confirmed WISPIT 2b.
The astrometry rules out a background source and marginally detects orbital motion of WISPIT 2c, which needs further follow-up observations for confirmation.

WISPIT 2 now becomes an analogue to PDS 70, offering a second laboratory for studying the formation and early evolution of a multi-planet system within its natal disk.
\end{abstract}



\section{Introduction} 
\label{sec:intro}

The detection of thousands of extrasolar planets with indirect detection methods indicates that planet formation is an efficient process that happens commonly around young stars \citep[see e.g.][]{Lissauer2023}.
At the same time, high resolution observations of planet-forming disks at multiple wavelengths have now revealed intricate sub-structures in more than a hundred systems \citep[see recent surveys by][]{Avenhaus2018, Andrews2018, Oberg2021, Garufi2024, Ginski2024}.

This indicates that planet-formation is indeed ongoing in many of these systems, yet the direct detection of embedded planets has proven challenging \citep[e.g. see discussion in][]{Milli2012, Currie2023}.
While there are mechanisms that can shape transitional disks without requiring the presence of planets \citep[][]{Marcus2015, Flock2017, Suriano2018, Kurtovic2018, Kuznetsova2022}, many of these systems provide strong indirect evidence for ongoing planet formation.
To date there are only a few such young, disk-bearing systems in which planets or strong candidates have been confirmed, such as PDS\,70 \citep{Keppler2018, Haffert2019}, AB\,Aur \citep{Currie2022}, and HD\,169142 \citep{Hammond2023}. 
It is the detection of these planets in formation that gives us key insights into the primary formation mechanisms that are at work in the earliest phases of system evolution, be it through core accretion \citep{Pollack1996} or gravitational instability \citep{Boss1997}.

Recently, the WISPIT\,2 system has been added to the short list of confirmed planet-bearing disks \citep{VanCapelleveen2025, Close2025}.
Uniquely among systems with a confirmed embedded planet, the planet-forming disk in the WISPIT\,2 system shows an extended multiple-ringed structure.
The primary star in the system is a young solar analogue \citep{VanCapelleveen2025} of 5.1$^{+2.4}_{-1.3}$ Myrs, based on stellar isochrones, which gives the intriguing possibility to observe the earliest formation history around a system similar to our own.
The embedded planet within the system is located between the innermost two rings at a semi-major axis of 57$^{+8}_{-3}$\,au and is estimated to have a mass of 4.9$^{+0.9}_{-0.6}$\,\mj{} \citep{VanCapelleveen2025}.
The  presence of up to four confirmed rings with intermediate gaps, as well as a central cavity in the system, gave rise to speculation that this could be a multi-planet system in formation, possibly comparable to PDS\,70, or a younger version of HR\,8799 \citep{Marois2008}.
Indeed \citet{Close2025} pointed out that a signal (referred to as CC1) appears to be present in their $L$- and $z'$-band observations at a projected separation of $\sim$15\,au within the central disk cavity.
In this letter we present new observations of this companion candidate using VLT/SPHERE \citep[Spectro-Polarimetric High-contrast Exoplanet REsearch; ][]{Beuzit2019} and VLTI/GRAVITY \citep{GRAVITY2017} confirming the presence of a second planet in the WISPIT\,2 system. 

\section{Observations}

In addition to the observations presented in \citet{VanCapelleveen2025}, we obtained new high-angular resolution data with both the VLT/SPHERE Infrared Dual-band Imager and Spectrograph \citep[IRDIS; ][]{Dohlen2008} and VLTI/GRAVITY, providing up to date $H$-band images and $K$-band interferometry observations, respectively. 

\subsection{GRAVITY Observations}

\begin{table*}[t]
\centering
\begin{tabular}{cccccccc}
\hline \hline
Target & UTC Date & Instrument & Setup & NEXP/NDIT/DIT & Airmass & $\tau_0$ (ms) & seeing (\arcsec)\\
\hline 
WISPIT 2 &2025-10-05& VLTI/GRAVITY & MEDIUM & 12/8/30s & 1.1-1.3 & 3.1-5.1  & 0.51 - 0.79 \\
HD 180595 (CAL) & 2025-10-05 &VLTI/GRAVITY & MEDIUM & 1/32/3s & 1.6 & 4.3  & 0.59 \\
WISPIT 2 & 2025-09-24 &VLT/SPHERE & $H$-band & 48/1/64s & $1.2\pm0.1$ &  $5.7\pm1.2$ & $0.8\pm0.2$\\
\hline
\end{tabular}
\caption{Log of the GRAVITY and SPHERE observations.}

\label{tab:log}
\end{table*}


The GRAVITY observations \citep{GRAVITYCollab2017} were carried out on the four Unit Telescopes (UT) on 2025-10-05.
The observations were executed in Service Mode (program ID: 115.29HG.001) with good-to-average atmospheric conditions, and a spectral resolution of $R\sim500$.
The log of the observations is included in Table \ref{tab:log}.
The GRAVITY observations were obtained in dual-field on-axis mode.
In this mode, the Fringe-Tracker (FT) fiber stabilises the fringes on the host star, and the Science Channel (SC) fiber alternates between the planet and the star \citep{GCollab2019}.
The pointing of the SC fiber was informed by the SPHERE $H$-band images and with an initial guess at $\Delta$RA, $\Delta$Dec = (-22.87 mas, -107.6 mas) from the host star.
These observations benefit from the GRAVITY+ extreme Adaptive Optics system \citep[GPAO; ][]{GPAO2025} to reduce the speckle noise close-in to the host star and maximise the flux injected in the planet, using the high-order correction with the Natural Guide Star sensor \citep{Bourdarot2024}.
The data were reduced using the ESO GRAVITY pipeline \citep{Lapeyrere2014}.
The observations were taken together with an interferometric calibrator in order to measure normalised visibilities.
We used the \texttt{exogravity} pipeline \footnote{\url{https://gitlab.obspm.fr/mnowak/exogravity}} to extract the relative astrometry and contrast spectrum of the companion relative to the host star.
We integrate the GRAVITY spectrum over the 2MASS $K_s$-band filter curve to obtain the $K_s$-band magnitude included in Table~\ref{table:phot+astrometry}.

\subsection{SPHERE $H$-band observations}
\label{sec:SPHERE-OBS}

To extract the properties of WISPIT 2c, we use the SPHERE $H$-band data observed on UTC date 2025-03-21 described in \citet{VanCapelleveen2025}. 
In addition, we obtained a new SPHERE/IRDIS $H$-band observation carried out as part of program 115.29HG.002 (PI C.~Ginski) on UTC date 2025-09-24 with a similar observation setup, see Table~\ref{tab:log}.
This observation includes a dedicated reference observation obtained with the star-hopping technique, alternating approximately every 10 minutes between WISPIT 2 and a reference star.
During the science sequence, we obtained 12 full polarimetric cycles, each comprising exposures at the 4 half-wave plate positions ($Q^+, Q^-, U^+, U^-$).
Each individual exposure had an integration time of 64 seconds, resulting in a total on-target DPI exposure of 51.2 minutes.
Over the full science sequence, the pupil-tracking configuration delivered a total parallactic angle rotation of 19.6 degrees. 
Between science cycles, we recorded 20 reference-star frames with an exposure time of 64 seconds. 

\begin{figure*}[]
   \centering
   \includegraphics[width=\hsize]{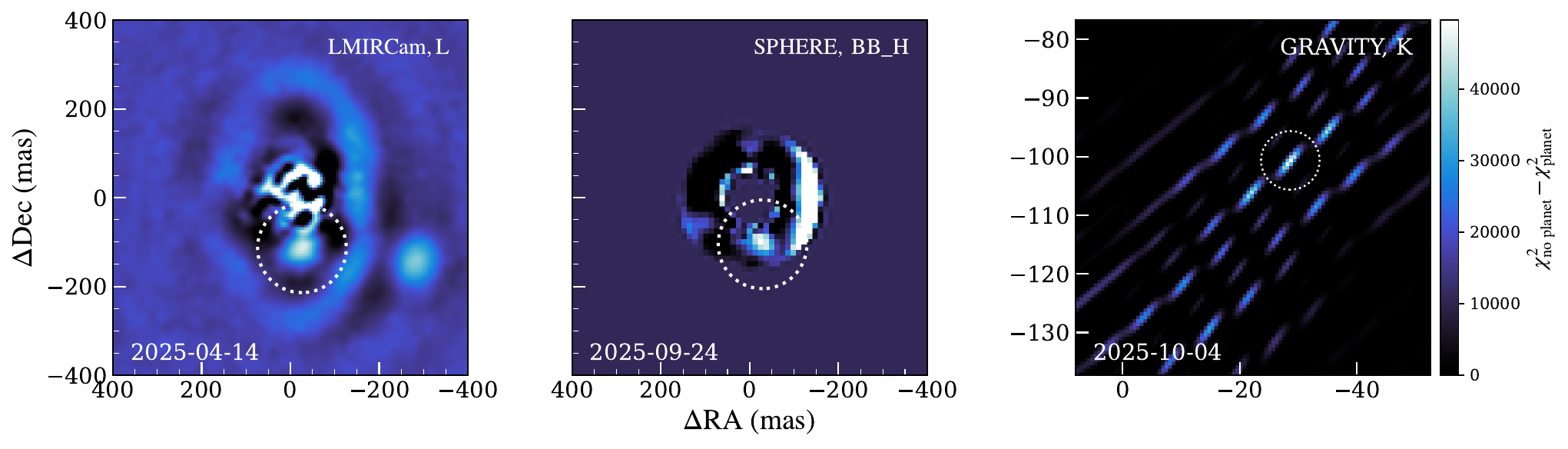}
      \caption{\textit{Left panel:} Original $L$-band detection of CC1 in \cite{Close2025}.
      \textit{Middle panel:} New $H$-band detection of WISPIT 2c following RDI.
      We show a confined annulus that was used for optimal stellar signal subtraction. 
      Note that the bright arch on the right side of the annulus is the forward scattering side of the circumstellar disk. 
      \textit{Right panel:} Detection map of WISPIT 2c planet with VLTI/GRAVITY on the night of 2025-10-04.
      The map shows the periodogram power maps of a point source companion as a function of angular offset from the star, where the companion location corresponds to the strongest peak.
      }

         \label{fig:detection_map}
   \end{figure*}

In this study we focus on the total intensity reduction of the SPHERE data and only used the polarised intensity imagery to confirm that there is no strong dust scattering signal present at the position of the investigated companion.
We extracted the photometry and astrometry of the companion (see Table~\ref{table:phot+astrometry}) using negative planet injection combined with a simplex minimisation technique, followed by Bayesian inference using Markov chain Monte Carlo \citep[MCMC; ][]{MacKay03} with 200 walkers and 10,000 steps, and repeated this method for fake positive injections at positions equidistantly distributed in polar space to extract the systematic errors of the method.
All steps are executed with \texttt{PynPoint} modules, and the centering precision of 2.5 mas of the star behind the coronagraph, as well as uncertainties in pixel scale, true North correction, and pupil offset, were included in the error budget, analogues to what is described in \cite{VanCapelleveen2025}.
However, instead of Angular Differential Imaging (ADI) we used Reference Differential Imaging (RDI) with Principal Component Analysis (PCA).
Given that the position of WISPIT 2c is on the inner working angle of the coronagraph, following the recommendation outlined in the SPHERE manual\footnote{SPHERE manuals: \url{https://www.eso.org/sci/facilities/paranal/instruments/sphere/doc.html}}, we apply a correction for the transmission profile of the coronagraph to both the science frames and the reference library.
As the observation from UTC date 2025-03-21 did not have a dedicated reference observation, we leveraged the $H$-band RDI library described in \citet{VanCapelleveen2025}.
For the observation of UTC date 2025-09-24 we used the dedicated reference-star observations.
We found the optimal number of principal components for RDI/PCA to be 12 and 14, respectively.
Due to the presence of the bright inner ring, the systematic errors of the method varied with position angle (PA).
To account for this we only included systematic errors similar to those near WISPIT 2c in our error budget; between a PA of 335 degrees and 15 degrees (to the north of WISPIT 2) and between a PA of 120 and 225 degrees (to the south of WISPIT 2).
The detection of WISPIT 2c after PCA/RDI in both observations is included in Fig.~\ref{fig:detection_map}.

\section{Nature of the companion}

\subsection{Re-detection of the companion with GRAVITY and SPHERE}

We re-detect the planetary companion WISPIT 2c with both GRAVITY and SPHERE (see Fig.~\ref{fig:detection_map}), providing independent confirmation of the source previously referred to as CC1 and reported by \citet{Close2025}.
The detection of embedded, self-luminous protoplanets is a particularly challenging task.
In the case of young planets embedded in a protoplanetary disk, it is necessary to distinguish the contribution of the scattered light of the disk from the thermal emission from the planet.
Substructures in the disk are usually of comparable scales to the Point Spread Function of 10m class telescopes.
Interferometry is intrinsically sensitive to the coherent emission of point-like sources, which is distinct from the extended background disk signal.
The GRAVITY data are key observational evidence for the detection of a new protoplanet in this work. 

We followed the approach of \citet{GCollab2019} and \citet{Wang2021} to model the coherent flux of the planet $V_{planet}(b,t,\lambda)$ measured with GRAVITY:
\begin{equation}
V_{planet}(b,t,\lambda) = C(\lambda) V_{star}(b,t,\lambda) \exp(i\Phi(b,t,\lambda))
\end{equation}

expressed as a function of $b$ the baseline, time $t$ and wavelength $\lambda$, with $C(\lambda)$ the contrast spectrum $C(\lambda)$ estimated from the \texttt{exoGRAVITY} pipeline, as described in \cite{Nowak2020}, and $V_{star}(\lambda)=F_{star}(\lambda)J_{star}(b,t,\lambda)$ the coherent flux of the star. The normalised visibility of the star $J_{star}(b,t,\lambda)$ was calibrated on a reference star HD 180595. We modelled the star spectrum $F_{star}(\lambda)$ using a \texttt{BT-Settl-CIFIST} stellar model (Fig.~\ref{SED}), as detailed in Appendix \ref{appendix:A}.

The phase $\Phi(b,t,\lambda)$ of the planet signal is defined as:

\begin{equation}
\Phi(b,t,\lambda)=-\frac{2\pi}{\lambda}(\Delta\text{RA}.u+\Delta\text{DEC}.v)
\end{equation}

wherein $(u,v)$ are the coordinates in the frequency domain, and ($\Delta\text{RA}$, $\Delta\text{Dec}$) the sky-coordinates of the planet relative to the star. The detection map using this point-source model is shown in Fig.~\ref{fig:detection_map}.  The detection map shows the periodogram  power maps $z(\Delta \text{RA}, \Delta \text{DEC})=\chi^2_{\text{no planet}}-\chi_{\text{planet}}^2(\Delta \text{RA}, \Delta \text{DEC})$ over the GRAVITY fiber field-of-view, as described in Eq. (B.6) of \citet{Nowak2020}. 
 The peak of the detection indicates a point-like source, with the side lobes characteristic of the uv-coverage of the UT interferometric array. 
 The planet is detected with a SNR$>$10 in each of the 12 exposures with GRAVITY.
 The best-fit is achieved at a location of $\Delta$RA, $\Delta$Dec = ($-29.07\pm0.024$ mas,$-101.35\pm0.038$ mas).
 The presence of a CPD in addition is discussed in Section \ref{sec:cpd}.

\subsection{Is it a bound planet?}
To ascertain the nature of the object labelled CC1 by \citet{Close2025}, we follow two independent lines of investigation.
We first determine if, with currently available astrometry, we can rule out that the object is a distant background star and secondly we test if the GRAVITY spectrum as well as the existing photometry are consistent with an object of planetary nature.\\
The GRAVITY $K$-band observation yields a relative astrometry of 105.44$\pm$0.04 mas and 196.00$\pm$0.01$\degree$.
However, the SPHERE $K_s$-band observation presented in \cite{VanCapelleveen2025} is unusable for astrometry extraction as the source lies too close to the coronagraph, influencing both the photometry and astrometry.
With no available published suppression profile of the $K_s$-band coronagraph, we cannot confidently extract either of these parameters.
The $L$-band astrometry presented in \cite{Close2025} is close to its resolution limit for its wavelength and therefore carries large error bars in both separation and position angle and is therefore not sufficiently constraining.
At such small separations we find that even if the PSF is misaligned by 0.5 of a pixel, this would change the position angle by approximately 2.8$\degree$, making the measurement highly sensitive to small systematics.
Although the $z'$-band data by \cite{Close2025} provide the longest baseline and smallest error bars, the source appears slightly extended and may include scattered light or CPD emission, which could bias the central position.
In addition to these literature values from \cite{Close2025}, we have extracted the astrometric position of the object from both SPHERE $H$-band observation epochs, i.e. the epoch taken in March of 2025 and originally published in \cite{VanCapelleveen2025} and the new SPHERE $H$-band observation taken in September of 2025 and presented in this study (see Appendix~\ref{appendix:SPHERE} Fig.~\ref{H-band_figs} for a side-by-side comparison). As we describe in section~\ref{sec:SPHERE-OBS} we applied a total intensity reference differential imaging reduction to both of these data sets which recovers WISPIT\,2c. Conversely to the SPHERE $K_s$-band epoch, a detailed coronagraphic throughput profile is available for the $H$-band observations in the SPHERE manual. We thus can reliably extract both astrometry and photometry of WISPIT\,2c.
However, because the position of WISPIT\,2c lies near the 50\% transmission of the coronagraph, the photometry is sensitive to the accuracy of the throughput correction. 
Uncertainty in the position of the star behind the coronagraph could lead to small variations in throughput correction, which has not been propagated into the photometric uncertainty of the companion.
While this may result in underestimated $H-$band magnitude uncertainties, this effect has been partially mitigated by averaging measurements from two independent epochs using two independent reference libraries.
Nevertheless, this does not address the possibility of a systematic magnitude offset resulting from over- or under-correction of the throughput if the transmission profile itself is not accurate.

\begin{figure}[h!]
   \centering
   \includegraphics[width=\hsize]{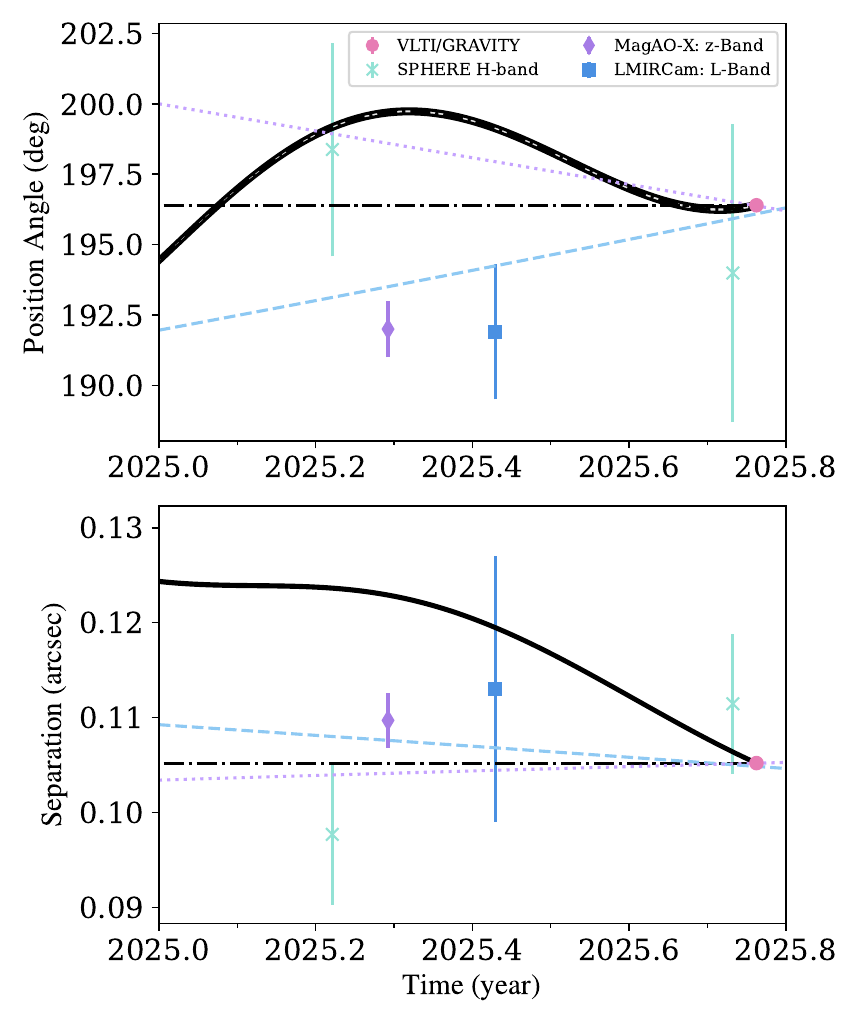}
      \caption{Relative astrometry of WISPIT\,2c to the primary star over time.
      We give the separation and position angle (measured from North over east).
      The purple dashed-line indicates plausible orbital motion for a prograde orbit, while the blue dashed-line indicates retrograde motion.
      Both orbits are in the plane of the circumstellar disk.
      The solid ``curved'' line indicates the expected behaviour for a distant background object.
      While the black dashed-line represents no relative change of motion in relation to the host star.}
         \label{fig:pm_plot}
   \end{figure}

We show all astrometric measurements in Fig~\ref{fig:pm_plot}. In the same figure we also illustrated the expected location of a non-moving background source. Due to the short time base-line the relative position angle of a gravitationally bound, co-moving source and a background object show considerable overlap. However, we do expect a strong change in separation for a background object that should not be present for a co-moving source. 
We find that in position angle the literature data from \cite{Close2025} in combination with our new GRAVITY observation appears to rule out a background object, while the SPHERE $H$-band epochs are consistent with both a background object and a co-moving bound planet. In separation the GRAVITY and SPHERE $H$-band data, as well as the literature $z'$-band data rule out a background object, while the $L$-band data uncertainties are too large to distinguish between both scenarios.
%
%
Given that the data points with the smallest uncertainties appear to rule out a non-moving background object, both in position angle and separation, we conclude that WISPIT\,2c is a bound object in the system. \\
However, we note that the object shows a measurable change in position angle and separation compared to the GRAVITY measurement.
This suggests that the object exhibits genuine orbital motion.
Assuming the planet lies in the disk plane (inclination fixed at 45$\degree$), we estimate the possible orbital solutions with \textit{orbitize!} \citep{Blunt2020} using an EMCEE MCMC sampler \citep{Foreman-Mackey2013}.
Based on the current astrometry we can explain the tentative change with Keplerian orbits, albeit with some inconsistency between the SPHERE $H$-band and literature $z'$ and $L$-band data points. While the former indicate orbital motion in the same direction as the outer planet WISPIT\,2b as reported in \cite{VanCapelleveen2025}, the latter exhibit a change in position angle which would put the orbit of 2c in retrograde relative to 2b. We illustrate both prograde and retrograde solutions in our proper motion analysis in Fig~\ref{fig:pm_plot}.
We note that a retrograde orbit would be extremely unusual and given the potential biases in the $z'$-band centroid and the limited precision of the $L$-band measurement, we find the pro-grade orbit indicated by the SPHERE $H$-band data to be the more likely scenario.
Additional high-precision astrometry will be required to confirm the orbital direction of WISPIT 2c.

With a distant background object ruled out by the astrometric analysis we inspected the $K$-band spectrum of WISPIT\,2c.
Fig.~\ref{spectrum} shows CO absorption in the GRAVITY $K$-band spectrum at 2.2935 $\text{\textmu m}$, and a positive slope from 2.15 - 2.25 $\text{\textmu m}$, both well-established signatures of young, low-gravity substellar objects \citep{Patience2012,Allers2013}.
%
%
Such CO absorption is expected to remain prominent in the atmospheres of young giant planets but is not produced at detectable levels in stellar photospheres beyond temperatures of $\sim$ 5300 K \citep{Cesetti2013}.
The photometric measurement in the $H$-band may follow the characteristic triangular continuum shape which arises from deep H$_2$O absorption both sides of the 1.65 $\text{\textmu m}$ peak \citep{Allers2013}, though our current data cannot confirm this as we cover this wavelength regime only with a single photometric point.

Taken together, both the astrometry as well as the spectroscopic data are strongly indicating that CC1 is indeed a second planet in the system which we will now refer to as WISPIT 2c.

\begin{table*}
\centering
\begin{tabular}{lcccccc}
\hline\hline
Instrument & UTC Date & Band & Magnitude & Separation (mas) & PA (deg) & Ref. \\
\hline
MagAO-X        & 2025-04-16 & $z'$ & $19.40^{+0.65}_{-0.26}$ & $109.7 \pm 2.9$   & $192.0 \pm 1.0$ & \citep{Close2025}\\
LBTI/LMIRcam   & 2025-06-05 & $L$ &$14.80^{+0.76}_{-0.43}$ & $113.0 \pm 14.0$  & $191.9 \pm 2.4$ & \citep{Close2025}\\
VLT/SPHERE  & 2025-03-21 & $H$ & $16.40 \pm 0.39$ & $97.69 \pm7.42$ & $198.38\pm 3.85$ & This work \\
VLT/SPHERE   & 2025-09-24 & $H$ & $16.25 \pm 0.29$ & $111.45 \pm 7.39$ & $193.99\pm5.25$ & This work \\
VLTI/GRAVITY   & 2025-10-05 & $K$ & $16.04 \pm 0.01$ & $105.44 \pm 0.03$  & $196.00 \pm 0.02$ & This work\\
\hline

\end{tabular}
\caption{Apparent magnitudes and relative astrometry of WISPIT~2c.}
\label{table:phot+astrometry}
\end{table*}

\subsection{Temperature and mass of WISPIT\,2c}

To assess the atmospheric properties of WISPIT 2c, we employed the spectral and photometric tool \texttt{species} \citep{Stolker2020b}, which is commonly used for the analysis of directly imaged planets and brown dwarfs.
We performed an atmospheric parameter inference by sampling self-consistent atmospheric model grids with the nested sampler \texttt{dynesty} \citep{Speagle2020} fitting the \texttt{Drift-Phoenix} \citep{Hauschildt1999,Baron2003,Woitke2003,Woitke2004,Helling2006,Helling2008a, Helling2008b,Witte2009, Witte2011} model.
We furthermore explore also several additional models, the results of which we describe in Appendix~\ref{Appendix:B}.

%
\texttt{Drift-Phoenix} is designed for the dust-rich atmospheres we expect of young embedded planets, and includes detailed cloud microphysics \citep{Rajpurohit2012}.
For the atmosphere constraint we consider the $z'$- and $L$-band photometric points by \citep{Close2025} as well as our new SPHERE $H$-band photometric measurement and GRAVITY $K$-band spectrum.
We discuss in Appendix~\ref{Appendix:B} how the recovered parameters are influenced if we use various sub-sets of these data points.
As we are not sensitive to variations in log \textit{g} (see Appendix~\ref{Appendix:B} Fig.~\ref{logg}), varying log \textit{g} between 3 and 5 does not significantly modify the inferred atmospheric parameters.
Taking an average of the available log \textit{g} values for young ($<50$ Myr) gas giants \citep{Schneider2011}, returns an average log \textit{g} of $\sim$3.9, which we take as a fixed value.
As WISPIT 2 is a solar type star we fix [Fe/H] at 0.0, similar to log \textit{g} we are not sensitive to changes in metallicity (see Appendix~\ref{Appendix:B} Fig.~\ref{metallicity}).

\begin{figure*}[t]
   \centering
   
   \includegraphics[width=\hsize]{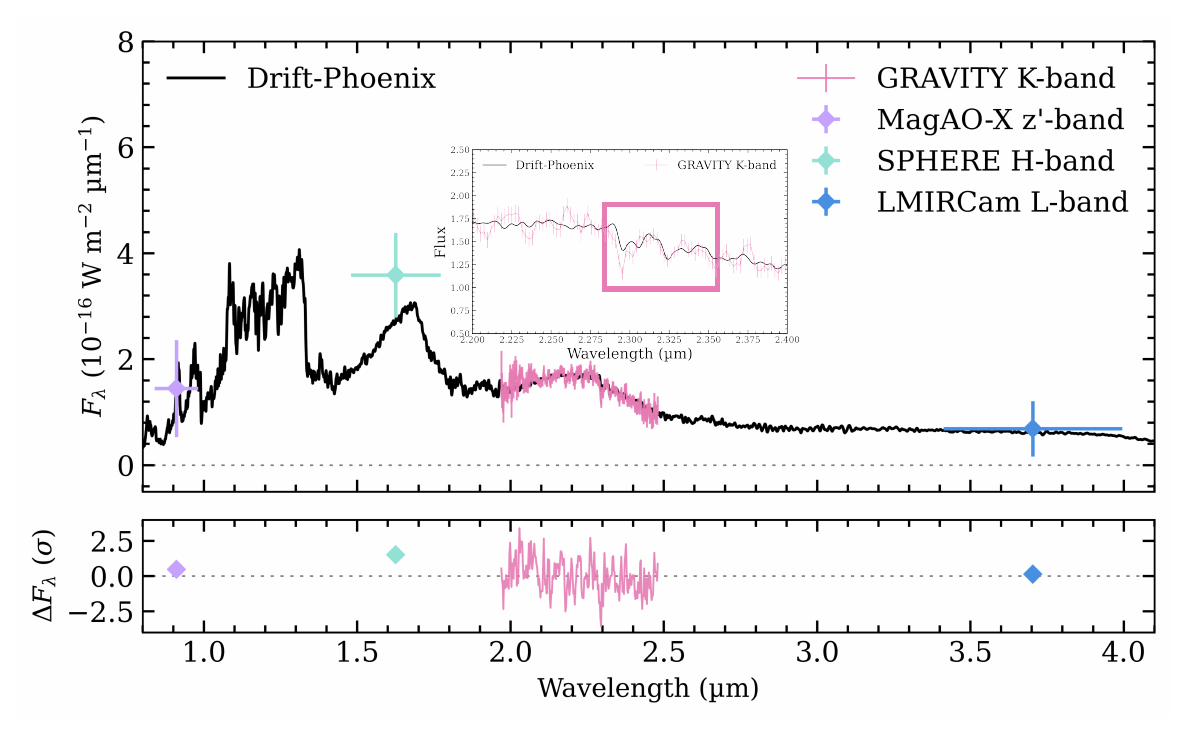}
      \caption{\texttt{Drift-Phoenix} spectral model overlaid with the $K$-band spectrum obtained with VLTI/GRAVITY, $H$-band VLT/SPHERE, $z'$-band MagAO-X, and $L$-band LBTI/LMIRCam photometry of WISPIT 2c. The modelled spectrum is based on max likelihood fitting of temperature and radius.
      Figure also contains zoomed region of 2.2-2.4$\text{\textmu m}$ to highlight CO absorption band-heads, marked by the pink box.}

      \label{spectrum}
   \end{figure*}

\begin{figure}[h!]
   \centering
   \includegraphics[width=\hsize]{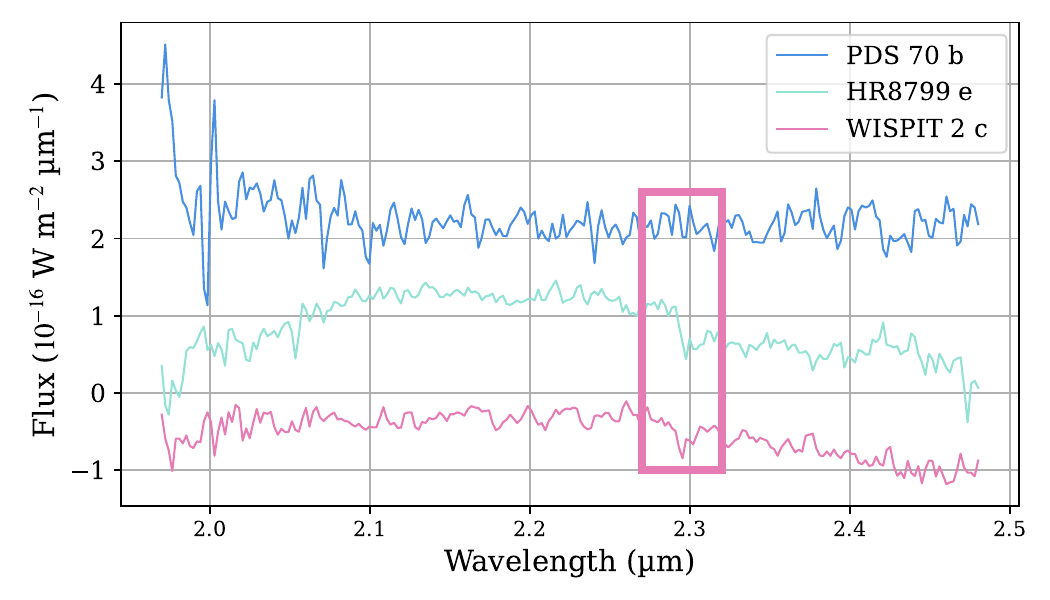}
      \caption{$K$-band GRAVITY spectra of PDS 70b and HR8799 e \citep{Kammerer2025}, and WISPIT 2c, offsets have been applied along the y-axis The CO feature (highlighted by the pink box) can be clearly seen in HR8799, and WISPIT 2c, but is lacking in PDS 70b. }
         \label{fig:CO_comparison}
   \end{figure}

Employing the \texttt{Drift-Phoenix} model, we derive max likelihood planet parameters of T$_{eff}$ = 1754$\pm$16 K, and R = 1.78$\pm0.03$ \rj{}, where the errors are purely statistical (see Appendix~\ref{Appendix:B}, Fig.~\ref{cornerplot_drift} for model distributions).
Since the pure statistical uncertainties are clearly underestimating the possible temperature and radius range for the planet, we considered the full set of explored models in Appendix~\ref{Appendix:B} to obtain a temperature range of 1500-2600 K, and a radius range of 0.91-2.20 \rj{}. \\
The distribution of temperatures appears bimodal, hence we separate the models into high and low temperature families, where more specific ranges can be seen in Table~\ref{planet_params}.
%
%
We note, however, that the ranges produced by the \texttt{BT-Dusty}, \texttt{Sonora-Bobcat}, \texttt{BT-Settl}, and \texttt{ATMO} models appear unphysical, as such a young object is expected to have an inflated radius (see for example \citealt{Spiegel2012}).
Given that the \texttt{Drift-Phoenix} model is a closer fit to our $K$-band spectrum with minimal residuals and produces more reasonable temperature and radius values, we consider these values more likely.
This conclusion is further strengthened by comparison of the $K_s-$band magnitude and $H-K_s$ colour against evolutionary isochrones, (see discussion below), which imply inflated radii larger than $1.6$ \rj{}.\\
We show the best fit \texttt{Drift-Phoenix} model and overlay our extracted GRAVITY $K$-band spectrum and $z'$-, $H$-, and $L$-band photometry in Fig.~\ref{spectrum}.
The model spectrum closely follows our extracted $K$-band spectrum, with a characteristic CO absorption at 2.2935 $\text{\textmu m}$ and overtones between 2.33-2.36 $\text{\textmu m}$, highlighted in the inset of Fig.~\ref{spectrum}.
The best fit results for the alternative models that we explored are shown in appendix~\ref{Appendix:B} (Fig.~\ref{spectrums}).
The \texttt{BT-Dusty}, \texttt{Sonora-Bobcat}, \texttt{BT-Settl}, and \texttt{ATMO} models converge toward high temperature ($\sim$2300-2600 K) solutions that imply unphysically small radii (0.91 - 1.07 \rj), inconsistent with expectations for a young, inflated object.
Given that our two best fitting models, \texttt{Drift-Phoenix} and \texttt{ExoRem}, i.e. the models with the smallest residuals, both include cloud physics, this may suggest that WISPIT 2c possesses a dusty atmosphere, given that the planet is still young, this is expected due to its low surface gravity \citep{Marley2011, Charnay2018}.\\
In addition to our model grid exploration, we compare the GRAVITY $K$-band spectrum of WISPIT\,2c with the spectra of PDS\,70b and HR8799\,e which are taken from the ExoGRAVITY $K$-band spectral library of gas giants and brown dwarfs presented in \citep{Kammerer2025}. These two objects were selected based on their similar absolute luminosity compared to WISPIT\,2c. The resulting comparative spectra are shown in Fig.~\ref{fig:CO_comparison}.
We note that the continuum shape and depth of CO absorption features of WISPIT\,2c and HR8799\,e are overall similar. While HR8799\,e is found to be slightly colder than WISPIT\,2c (1100-1400\,K, \citealt{Nasedkin2024}), it confirms that our spectrum is consistent with a young gas-giant planet. Conversely the spectrum of PDS\,70b, a planet still located within its' birth disk, appears almost flat with no strong CO feature. \cite{Wang2021} discuss that the spectrum is most likely explained by a planetary atmosphere with significant amounts of dust, the nature of which is uncertain. The difference in spectrum between PDS\,70b and WISPIT\,2c may then indicate that the former is still strongly embedded, while this is not true for the latter.\\

\begin{figure}[h!]
   \centering
   \includegraphics[width=\hsize]{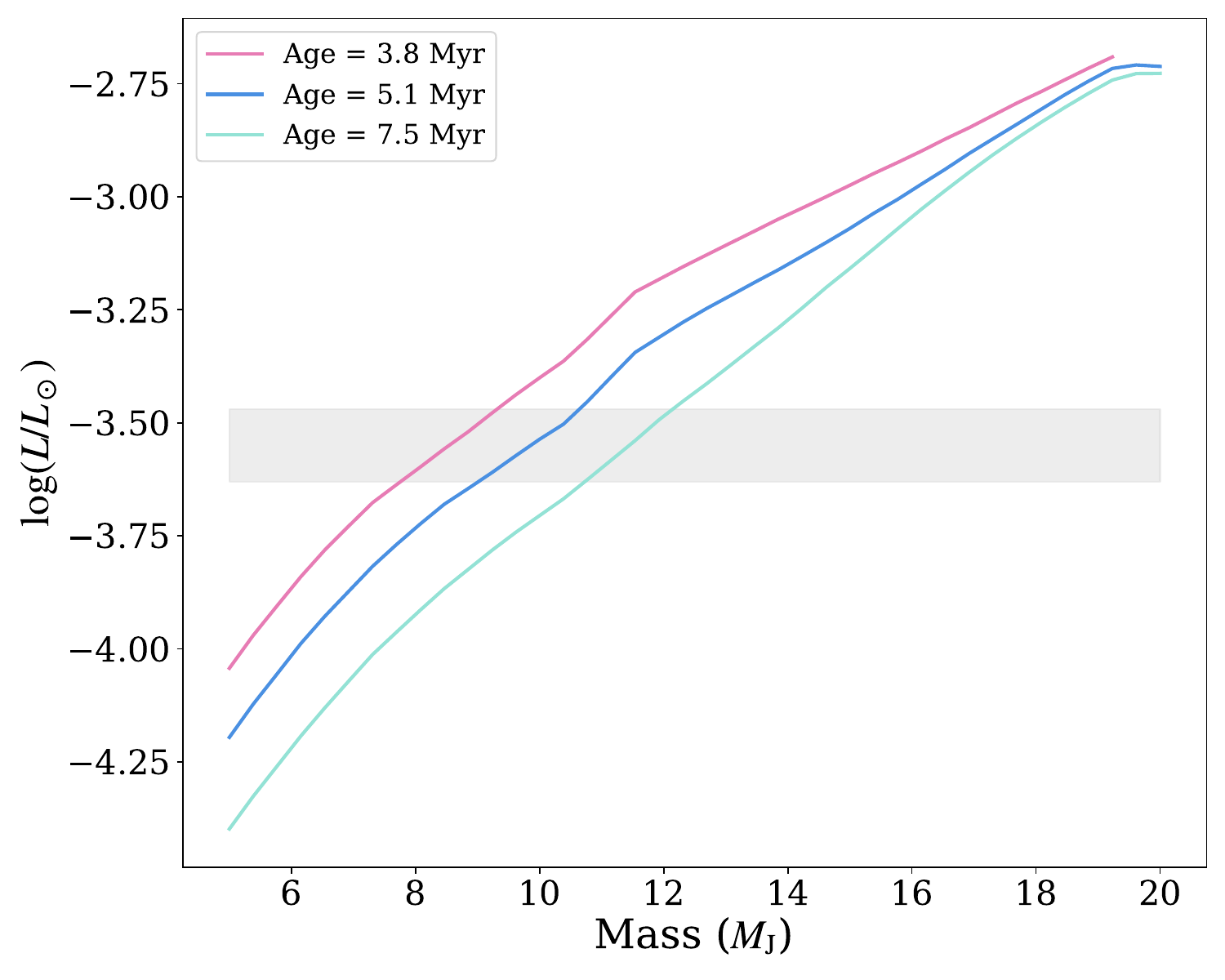}
      \caption{\texttt{Sonora-Diamondback} luminosity-mass isochrones at varying ages, measured luminosities from the atmospheric model fitting are shown by the grey shaded region.} 
        \label{Isochrones}
\end{figure}

Using the luminosity derived from our radius and temperature range, we estimate the mass of WISPIT 2c by comparison to evolutionary model isochrones.
To determine the sensitivity of our mass estimate to assumptions of age and evolutionary models, we computed isochrones for WISPIT 2c using a range of representative ages and atmosphere grids.
In Fig.~\ref{Isochrones}, we show isochrones at 3.8, 5.1, and 7.5 Myr (the system age of 5.1 Myr and its associated errors reported in \citealt{VanCapelleveen2025}).
For each age we compare the predicted luminosity-mass relations with our measured luminosity, including its uncertainty range, taken from our posterior samples.
Across this age span, we infer a corresponding mass range of approximately 8-12 \mj{}, consistent with the values previously reported for CC1 by \citet{Close2025} of ~8-10 \mj{}.
We tested how robust this estimate is against the usage of different model isochrones and found a mass range of roughly 9-11 \mj{}, i.e. smaller than the range given by the uncertainty of the system age. 
Based on these mass ranges, it appears likely that WISPIT\,2c sits comfortably in a range of 8-12 \mj{}. 
However, we note that our atmospheric grid models infer a broader range of masses of 3-16 \mj{}.
Since mass is not a fitted parameter in these models but is instead derived from the fixed 
log \textit{g} and inferred radius, we do not consider these to be robust mass estimates as our data is not sensitive to small changes in log \textit{g} as mentioned previously.

\begin{figure}[h!]
   \centering
   \includegraphics[width=\hsize]{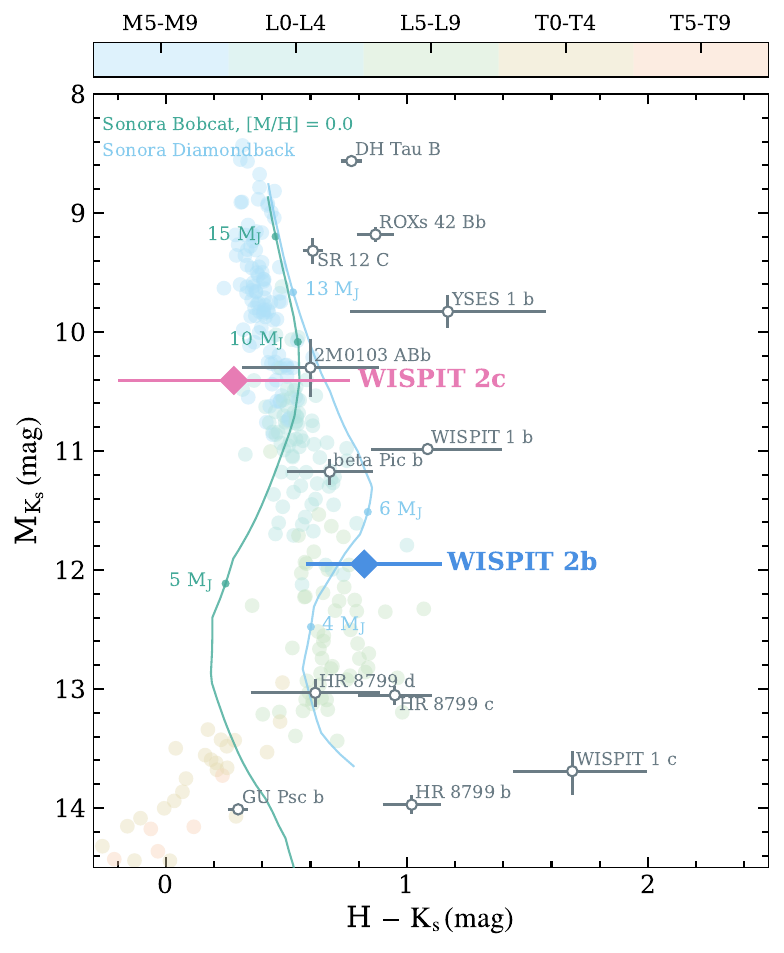}
    \caption{Colour-magnitude diagram of the WISPIT~2 system, along with confirmed planetary companions and field brown dwarfs of various spectral types. The teal and cyan tracks show 5.1~Myr \texttt{Sonora-Bobcat} (cloud-free) and \texttt{Sonora-Diamondback} (hybrid clouds, sedimentation efficiency $f_\mathrm{sed}=2$) evolutionary isochrones, respectively.}
         \label{fig:cmd}
   \end{figure}

In addition to our spectral analysis, the photometry of WISPIT\,2c is shown in a colour-magnitude diagram in Fig.~\ref{fig:cmd}, along with that of WISPIT\,2b and other known planets with available $H$- and $K_s$-band magnitudes.
Comparison against 5.1~Myr \texttt{Sonora-Bobcat} \citep{Marley2021} and \texttt{Sonora-Diamondback} \citep{Morly2024} isochrone tracks confirm that WISPIT\,2c is a planetary-mass companion, and that the $H-K_s$ $1\sigma$ upper bound is consistent with a cloudy model, in agreement with the best-fit atmosphere models to the combined spectrum and photometry.
We note that the lower temperature and inflated radius are also preferred from comparison against these evolutionary isochrones.

\begin{table}[h!]
\centering
\begin{tabular}{lccc}
\hline \hline
Planet parameters & High T & Low T\\
\hline
$\mathrm{T_{eff}}$ K & 2300-2600 & 1500-1750 \\
$\log g$ & $3.9$\footnote{Note the exception of the \texttt{BT-Dusty} model where log \textit{g} is 4.5.} & $3.9$\\
$[Fe/H]$ & 0.0 & 0.0\\
$\mathrm{R_p/R_{Jup}} $ & 0.91-1.07  & 1.78-2.20  \\
$\log \mathrm{L_p/L_{\odot}} $ & (-3.47)-(-3.53) & (-3.55)-(-3.63) \\
$\mathrm{M_p/M_{Jup}} $ & 3-14 & 10-16  \\
$\mathrm{M_p/M_{Jup}} $ & \multicolumn{2}{c}{8-12\footnote{Mass range derived from luminosity-mass isochrones.}}\\
\hline

\end{tabular}
\caption{Ranges of physical parameters of WISPIT 2c based on the atmospheric model fitting from the models presented in Appendix~\ref{Appendix:B}, were we separate based on high and low temperature families. Note: The ranges are based on the highest and lowest most likelihood values of each model, and we do not report statistical errors as they are negligible.}
\label{planet_params}

\end{table}

\section{Discussion}

\subsection{Comparison with PDS 70}

The young T Tauri star PDS 70 \citep{Gregorio-Hetem1992} once acted as a lone candle in the dark for early planet formation studies, owing to its two confirmed planets, PDS 70b \citep{Keppler2018} and PDS 70c \citep{Haffert2019}.
The confirmation of two planets in the young WISPIT 2 system now provides a valuable second comparison point, helping to narrow the observational gap in our understanding of how giant planets convene within their natal disks.
Although the sample of directly imaged, still embedded protoplanets remains too small for robust statistical analyses, it is still informative to compare the orbital architectures of WISPIT 2 and PDS 70.

PDS 70 exhibits a well-defined inner cavity, host to two planets, measuring roughly 70 au \citep{Hashimoto2012,Dong2012,Hashimoto2015,Keppler2019} similar to that of WISPIT 2's prominent 60 au gap, containing one planet, with a second planet sitting beyond the innermost ring.
Both systems show strong signs of being shaped by their respective planets.
Despite the more extended disk in WISPIT 2, the confirmed planets in both systems occupy a broadly similar radial regime, approximately 57 and 14 au for WISPIT 2b and c, and 21 and 35 au for PDS 70b and c \citep{Haffert2019}.
The lack of a dust ring between the planets in PDS 70 is likely due to the close separation between its respective planets.
These planets likely would have accreted material from both sides of the ring early in their formation stages, thus depleting the dust ring and forming the large cavity that now harbours these planets.
While the planets in WISPIT 2 are at a much larger separation and therefore are unable to efficiently clear the material between them, allowing the dust ring to remain intact.
Recent astrometric fits by \citet{Trevascus} place upper mass limits of 4.9 \mj{} and 13.6 \mj{} for PDS 70b and c respectively, comparable to the masses of WISPIT 2b and c.
These broadly similar radial and mass ranges may imply the existence of a ``Goldilocks zone'' for early giant planet formation, where conditions in both systems appear to be conducive to generating giant multi-planet architectures.
It is likely that the planets in both WISPIT 2 and PDS 70 formed in-situ  \citep[see ][]{VanCapelleveen2025,Perotti2023}, implying their current locations trace the environments that supported their formation rather than the endpoints of substantial inward or outward migration.

\subsection{Non-detection of H$\alpha$ and CPD}
\label{sec:cpd}

Despite its larger mass compared to the previously detected outer planet WISPIT\,2b, there was no significant H$\alpha$-emission detected at the position of WISPIT\,2c \citep{Close2025}.
Given that accretion generally scales with mass \citep[see e.g. ][]{Ercolano2014, Manara2023} this indicates that either the accretion on the object is highly variable, or that the object is veiled by circumplanetary dust, which effectively blocks the optical wavelength emission.
As we only have a single epoch of H$\alpha$ imaging of the system to date, we are unable to verify if the accretion signatures of the planet might be variable.
However, from large spectroscopic surveys we know this commonly to be the case for young stars \citep[e.g. ][]{Manara2021}.
Given that the transport mechanisms within the disk for accretion onto the central star or a surrounding planet are similar, it is reasonable to assume that this is indeed the case also for the accretion activity of the planet.
This indeed has been recently confirmed for the planets in the  PDS\,70 system \citep{Close2025a, Zhou2025}.

On the other hand \citet{Close2025} point out that the $z'$-band measurement appears to be tentatively extended, consistent with a CPD signal within the Hill radius of a massive planet.
%
%
Conversely, our preliminary GRAVITY observation rule out the presence of a compact CPD at scales of $\sim 0.25$ au.
This may indicate that if a CPD is indeed present, then the dust grain sizes within the disk are on a sub-micron scale, so that they efficiently scatter light in the optical $z'$-band but do not significantly contribute to the signal in the $K$-band.
Alternatively this could indicate that there is some larger scale dusty envelope present at the position of WISPIT\,2c, in fact very similar to what has been observed for PDS\,70c \citep{Stolker2020}.
This envelope could then be detected in $z'$-band but resolved out by the interferometric GRAVITY observation.
The presence of such an envelope could then also explain the non-detection at the H$\alpha$ wavelength.

\subsection{Planet orbital eccentricity and multiplicity}

The confirmation of a second planet in the WISPIT 2 system aligns with expectations based on its dynamical behaviour.
The previously detected planet in the system, WISPIT 2b, exhibits an extremely low eccentricity ($e\leq0.2$), consistent with trends observed in multi-planet systems.
Studies by \citet{Xie2016} found that from a sample of \textit{Kepler} planets, the average eccentricity for single planets is approximately 0.3, while systems with multiple planets tend to have significantly lower average eccentricities of 0.04.
Similarly, \citet{Limbach2015} reported a strong inverse correlation between planet multiplicity and orbital eccentricity. 
Follow-up observations will be key to constrain the orbits of both planets in the system, to establish solid links between orbital architecture and planet multiplicity.


\section{Conclusions}

Through combined photometric and spectroscopic analysis, we confirm the presence of an additional planetary mass companion in the WISPIT 2 system.
The CO band-head detected in the GRAVITY $K$-band spectrum provides strong evidence for a planetary object, which is twice as massive as WISPIT 2b and significantly closer to the host star, with a mass range of 8-12 \mj{} and a radial separation of 14 au.
While WISPIT 2b showed strong signs of H${\alpha}$ emission, there is an absence of significantly detectable H${\alpha}$ emission from WISPIT 2c, which may indicate strong variability or dust veiling.
With current available astrometry, the motion of the planet appears inconsistent with that of a distant background object, but further high-precision astrometric observations are needed to constrain its orbital motion.
With this new detection, WISPIT 2 becomes only the second system (after PDS 70) to host multiple directly imaged young giant planets in formation, making it a prime target for follow-up observations with the upcoming Extremely Large Telescope (ELT).
This offers a rare opportunity to probe how early system architectures emerge, 
Although still speculative, the comparable orbital separations in both systems may hint at a ``Goldilocks zone'' for giant planet formation in young disks.
While the available data remain limited, these results bring us one step closer to making direct connections between the initial conditions of planet formation and the final architectures of planetary systems. \\





\FloatBarrier
\begin{acknowledgements}
    The authors would like to thank an anonymous referee for comments that significantly improved the clarity of the article.
    
    CL would like to acknowledge support from the COST Action CA22133 PLANETS.
    
    The authors would like to thank Ilya Ilyin for fruitful discussion of their results.
    
    We thank the Fundação para a Ciência e Tecnologia (FCT), Portugal, for the financial support to the Center for Astrophysics and Gravitation (CENTRA/IST/ULisboa) through grant No. UID/\allowbreak PRR/\allowbreak 00099/\allowbreak 2025 (\url{https://doi.org/10.54499/UID/PRR/00099/2025}) and grant No. UID/00099/2025 (\url{https://doi.org/10.54499/UID/00099/2025}).

    SFH and CS acknowledge support through UK Research and Innovation (UKRI) under the UK government’s Horizon Europe Funding Guarantee (EP/Z533920/1, selected in the 2023 ERC Advanced Grant round). SFH and JL acknowledge support through an STFC Small Award (ST/Y001656/1).

    Part of this research was carried out in part at the Jet Propulsion Laboratory, California Institute of Technology, under a contract with the National Aeronautics and Space Administration (80NM0018D0004).

    Based on observations collected at the European Southern Observatory under ESO programmes 115.29HG.001 and 115.29HG.002.

\end{acknowledgements}

%
\bibliographystyle{aasjournal}
\bibliography{biblio}







   
  



\appendix





\section{Flux calibration of the GRAVITY data}
\label{appendix:A}
\begin{figure*}[h!]
   \centering
   \includegraphics[width=\hsize]{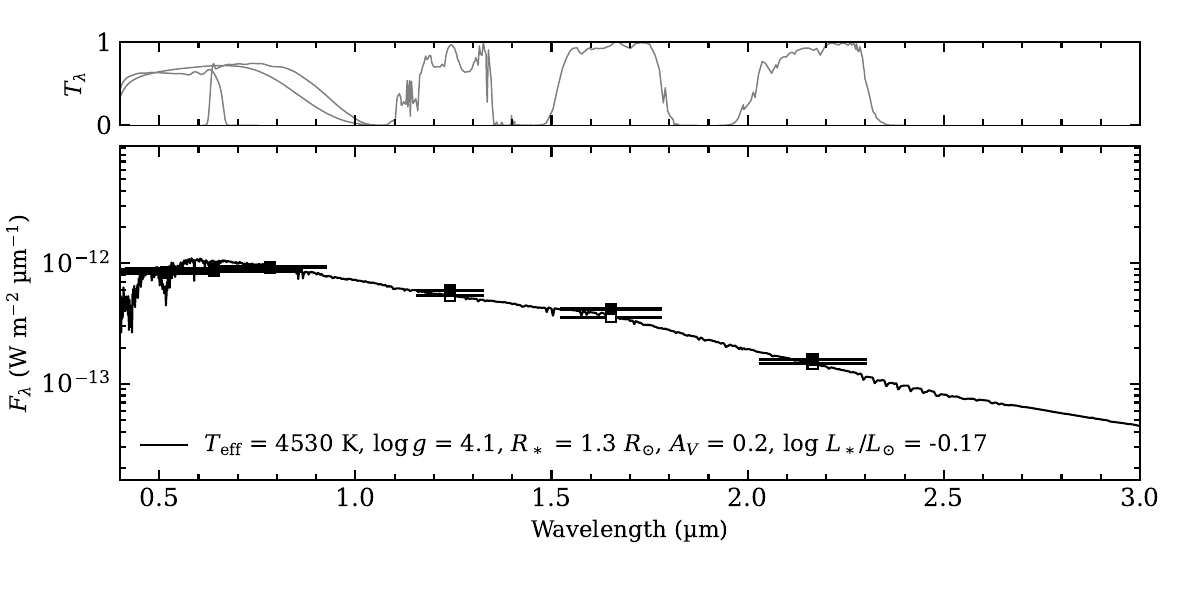}
      \caption{Fitted stellar SED of WISPIT 2 used for the absolute flux calibration of the GRAVITY data.}
         \label{SED}
   \end{figure*}

The SED from the star was obtained by fitting the photometry point from GAIA DR3 \citep{Gaia2023} and 2MASS \citep{Cutri2003}, used in \citet{VanCapelleveen2025}.
The absolute flux of the star was obtained with a synthetic spectrum from the \texttt{species}\footnote{\url{https://species.readthedocs.io/}} package, using a \texttt{BT-Settl-CIFIST} stellar model.
The result of the fit of the SED is shown in Fig.~\ref{SED}.
Finally, the absolute flux of the planet spectrum is obtained by multiplying the contrast spectrum extracted from the \texttt{exogravity} pipeline with the synthetic spectrum of the star.

\section{Comparison of the two SPHERE observation epochs}
\label{appendix:SPHERE}

\begin{figure*}[h!]
   \centering
   \includegraphics[width=0.8\hsize]{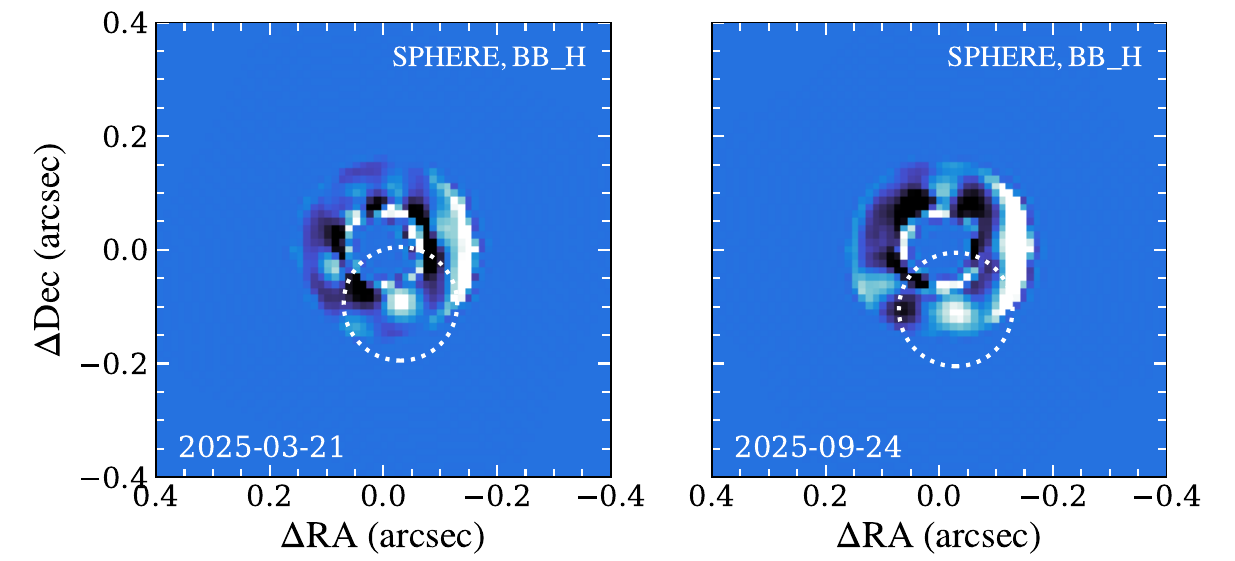}
      \caption{SPHERE $H$-band observations of the new planet WISPIT 2c. Both observation epochs clearly recover the planet (marked with the dotted circle). A mask is applied to exclude the majority of disk signal from the extraction and prevent over-subtraction. Both images utilise a linear colour map.}
         \label{H-band_figs}
   \end{figure*}

We show the RDI reduced (total intensity) SPHERE observations taken in March and September 2025 in Fig.~\ref{H-band_figs}. Both were taken with similar instrument settings and in the $H$-band filter. The main difference between the two epochs is the applied reference star data. While for the September 2025 epoch a dedicated reference star was observed in ``star-hopping mode", i.e. interleaved with the science observation, this was not the case for the March 2025 data. For this data set instead a reference library was utilised. Both observation epochs recover the companion at high significance.

\section{RGB image of the WISPIT 2 system}
\label{appendix:RGB}

\begin{figure*}[h!]
   \centering
   \includegraphics[width=0.8\hsize]{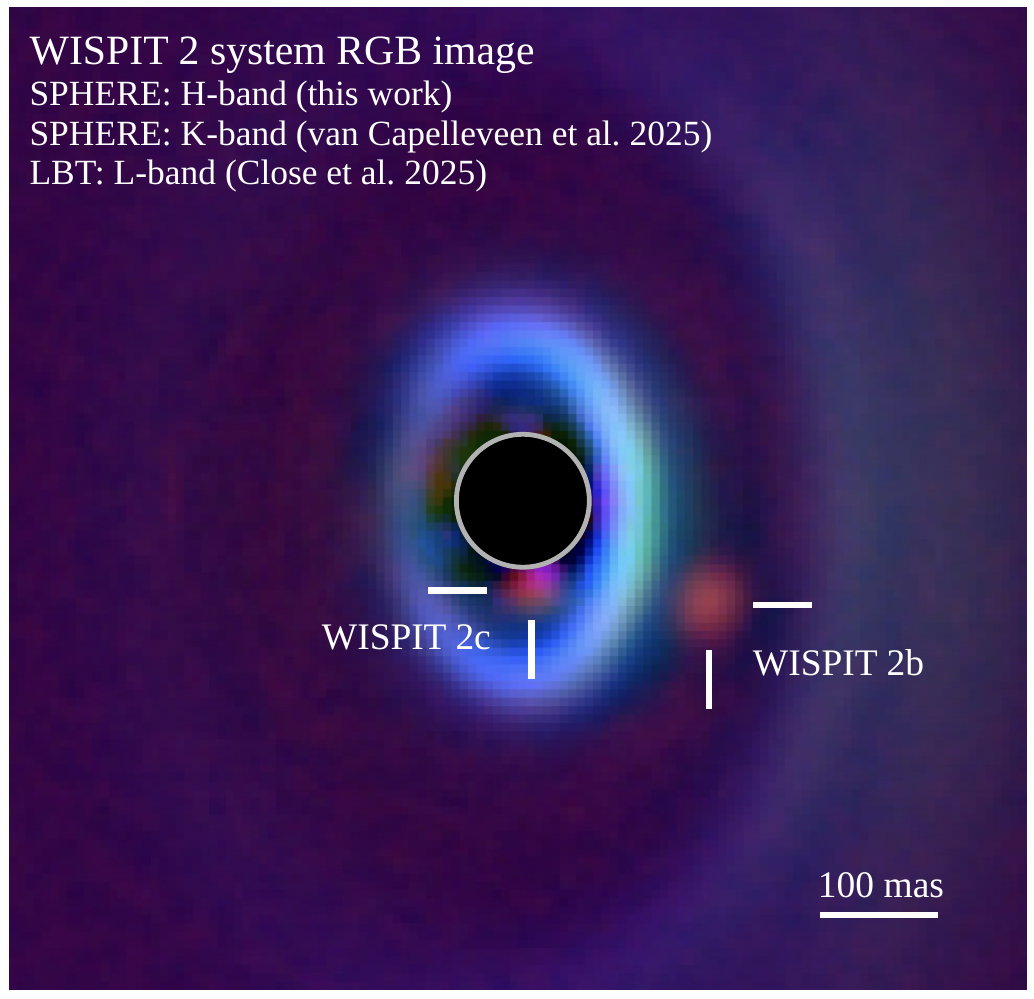}
      \caption{RGB image of WISPIT 2 using the existing $L$- and $K$-band data by \cite{Close2025} and \cite{VanCapelleveen2025}, as well as our new $H$-band data. We mark the positions of the known planet WISPIT 2b as well as the new inner planet WISPIT 2c.}
         \label{RGB-figure}
   \end{figure*}

In Fig.~\ref{RGB-figure} we show an RGB image of the WISPIT 2 system using existing literature data as well as our new SPHERE $H$-band observations. The red channel in the image is comprised of the ADI treated $L$-band data from \cite{Close2025}, while the green channel uses the RDI treated SPHERE $K$-band data first presented in \cite{VanCapelleveen2025}. The blue channel uses our newly obtained SPHERE $H$-band data. Here we use itself a composite of the RDI treated annulus confined data that recovers the planet WISPIT 2c and which is shown in Fig.~\ref{fig:detection_map}, and a polarisation image showing the disk scattered light created from the same data set (but not sensitive to the planet). Each channel was normalised such that the disk has a similar brightness towards is forward scattering peak. The brightness of each channel is scaled with the square root of the pixel values, which provides good contrast between planet signal and noise. We note that this image should be primarily considered for illustration of the system architecture and is not meant for an accurate colour analysis of the disk or the planet signal.

\section{Full nested sampling inference of planet atmosphere parameters}
\label{Appendix:B}
We performed a full atmospheric parameter inference of WISPIT 2c using the \texttt{species} tool \citep{Stolker2020b}.
The analysis combines both the medium resolution $K$-band spectrum extracted from the GRAVITY observations, using the standard deviation errors of the spectrum, and the $z'$-, $H$- and $L$-band photometry.

%

We use the \texttt{Drift-Phoenix} \citep{Hauschildt1999,Baron2003,Woitke2003,Woitke2004,Helling2006,Helling2008a, Helling2008b,Witte2009, Witte2011}, \texttt{Sonora-Diamondback} \citep{Morly2024}, \texttt{ExoRem} \citep{Baudino2015,Charnay2018}, \texttt{BT-Dusty} \citep{Allard2012}, \texttt{Sonora-Bobcat} \citep{Marley2021}, \texttt{BT-Settl} \citep{Allard2014}, and \texttt{ATMO} \citep{Phillips2020} models to explore a variety physical properties for the planetary atmosphere.
The model grids cover a wavelength range of 0.5-10 $\text{\textmu m}$ with a spectral resolution of 500 (consistent with that of the GRAVITY instrument) and 1000 live points.
We adopted uniform priors with reasonable ranges of 1500 - 3000 K and 0.5 - 3.5 R$_{J}$.
%
We fix the surface gravity $\log g$ at 3.9 for all models (average of reported values for young gas giants), with the exception \texttt{BT-Dusty}, where it was fixed at 4.5, as this is the minimum value accepted by the model grid.

\begin{figure*}[h!]
   \centering
   \includegraphics[width=\hsize]{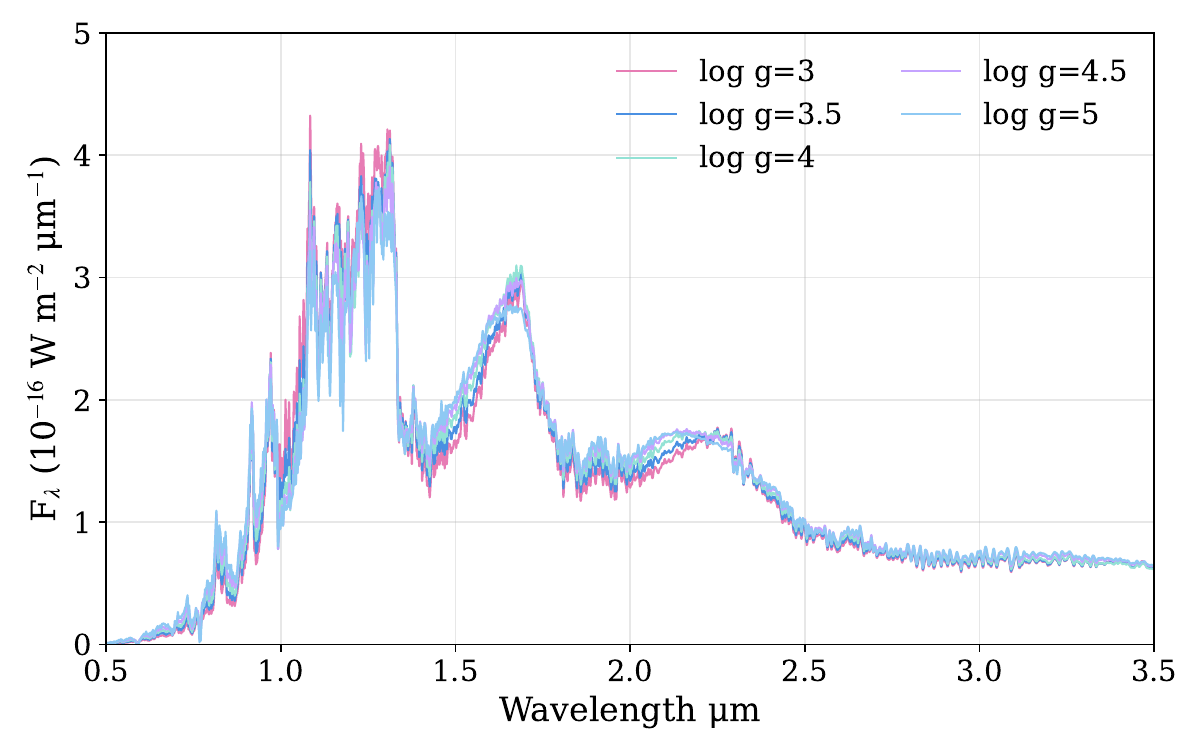}
      \caption{\texttt{Drift-Phoenix} model spectrum of the WISPIT 2c atmosphere with varying log \textit{g}, at a fixed temperature of 1750 K and a radius of 1.78 \rj{}}.
         \label{logg}
   \end{figure*}

\begin{figure*}[h!]
   \centering
   \includegraphics[width=\hsize]{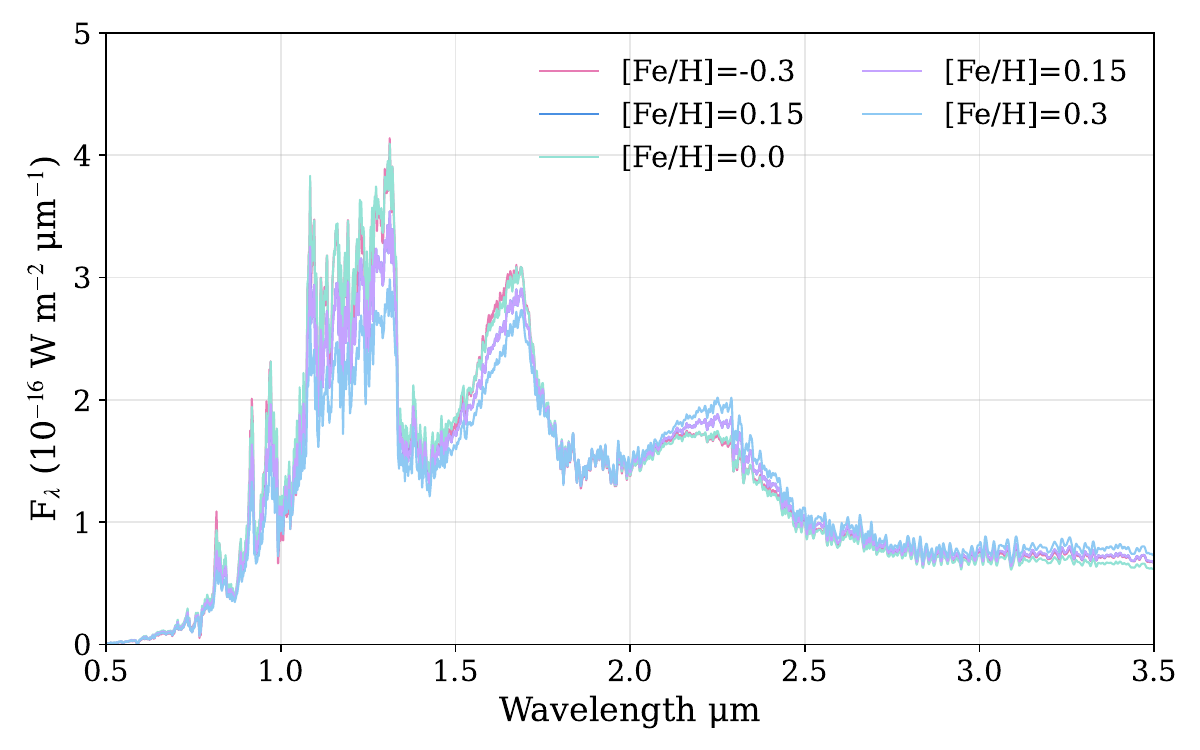}
      \caption{\texttt{Drift-Phoenix} model spectrum of the WISPIT 2c atmosphere with varying metallicity, at a fixed temperature of 1750 K and a radius of 1.78 \rj{}}.
         \label{metallicity}
   \end{figure*}

\begin{figure*}[h!]
   \centering
   \includegraphics[width=\hsize]{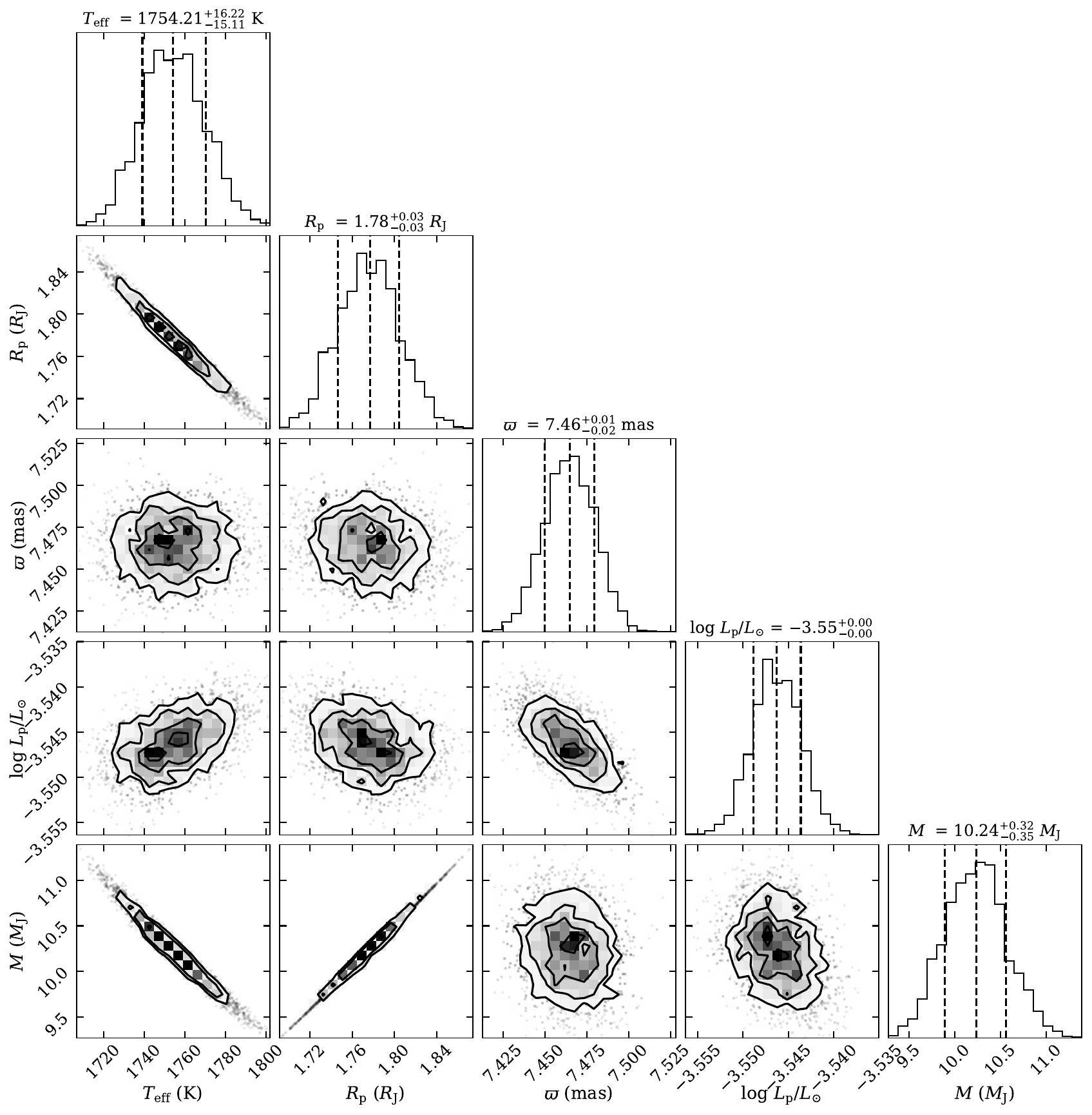}
      \caption{Posterior distributions for WISPIT 2c from the nested-sampling using \texttt{Drift-Phoenix}, with median values shown and $\pm 1\sigma$.}
         \label{cornerplot_drift}
   \end{figure*}

\begin{figure*}
\centering

\gridline{
\fig{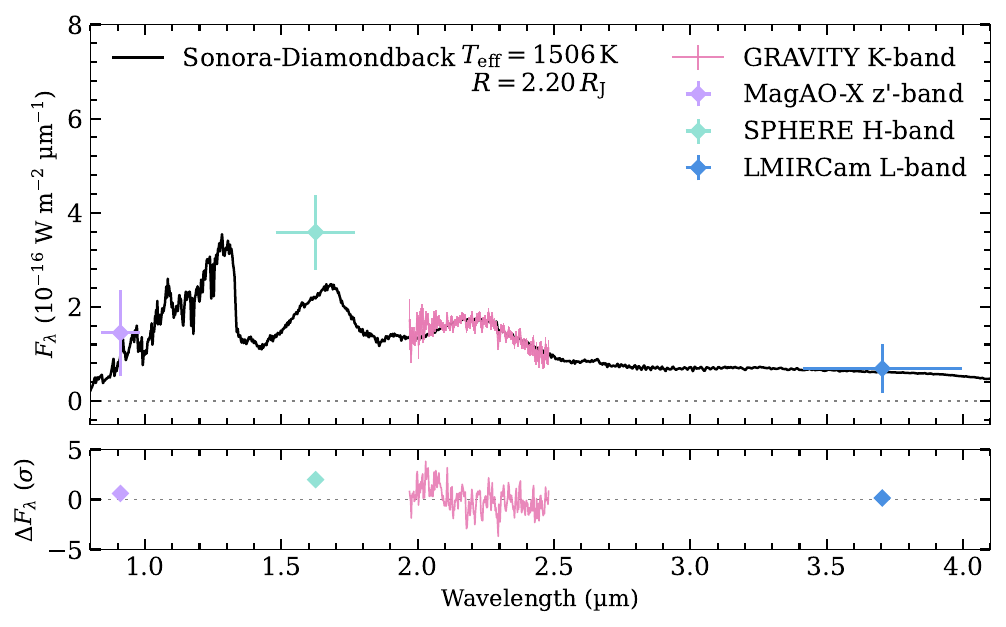}{0.48\textwidth}{(a)}
\fig{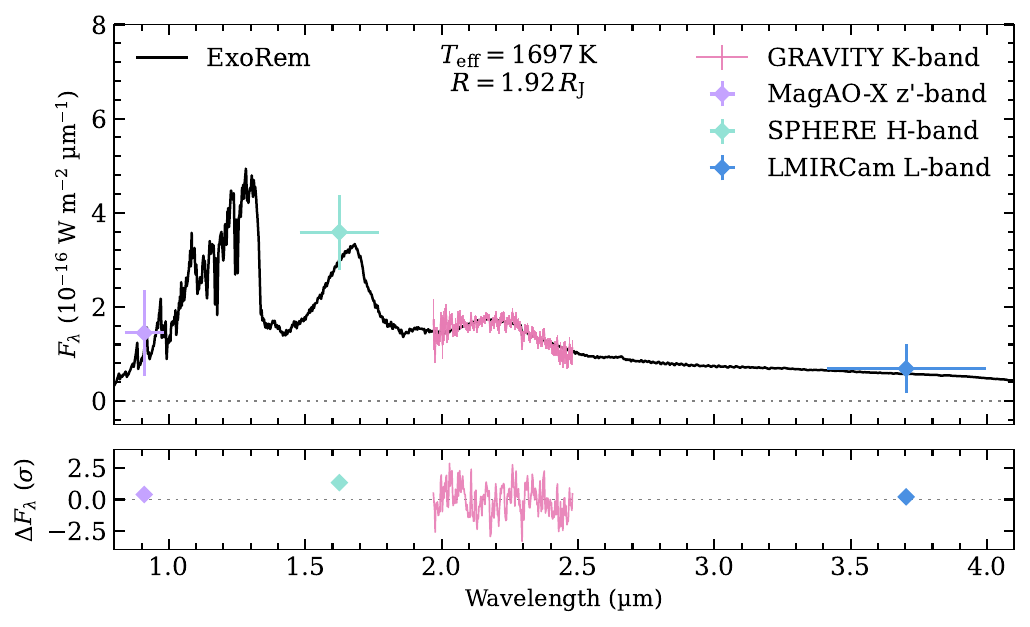}{0.48\textwidth}{(b)}
}

\gridline{
\fig{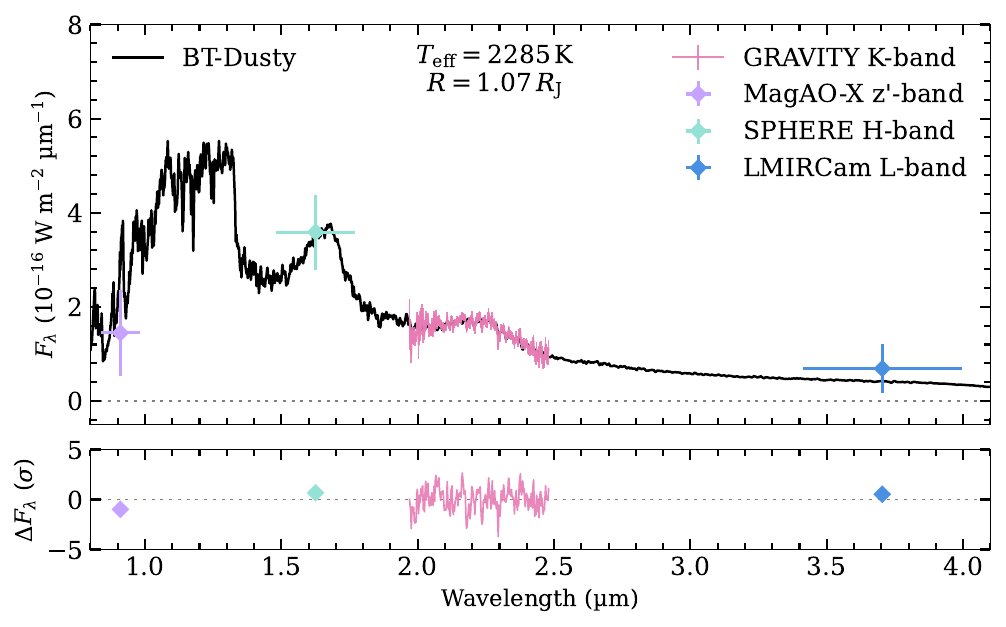}{0.48\textwidth}{(c)}
\fig{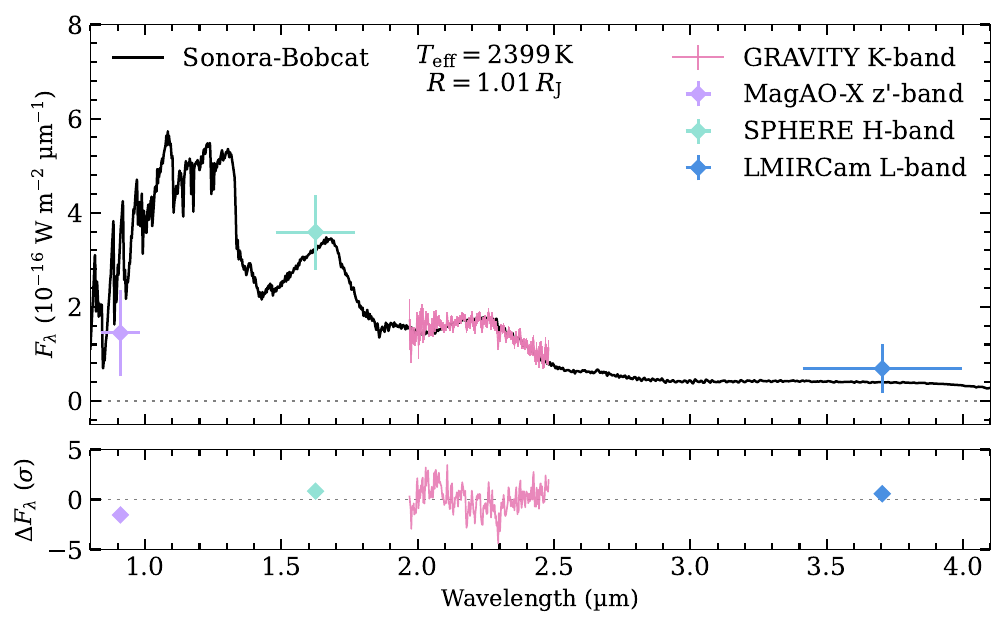}{0.48\textwidth}{(d)}
}

\gridline{
\fig{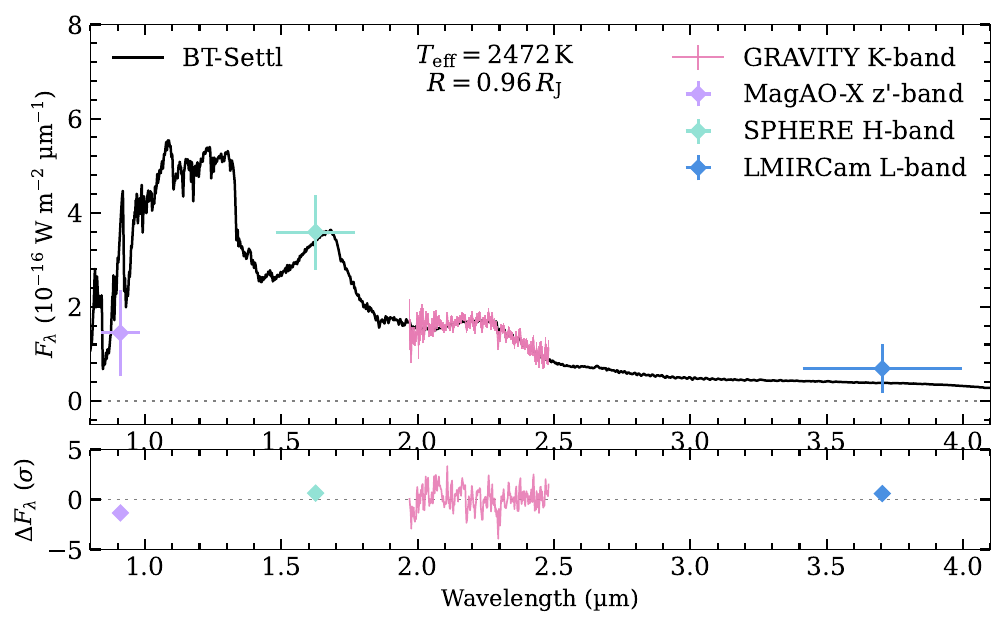}{0.48\textwidth}{(e)}
\fig{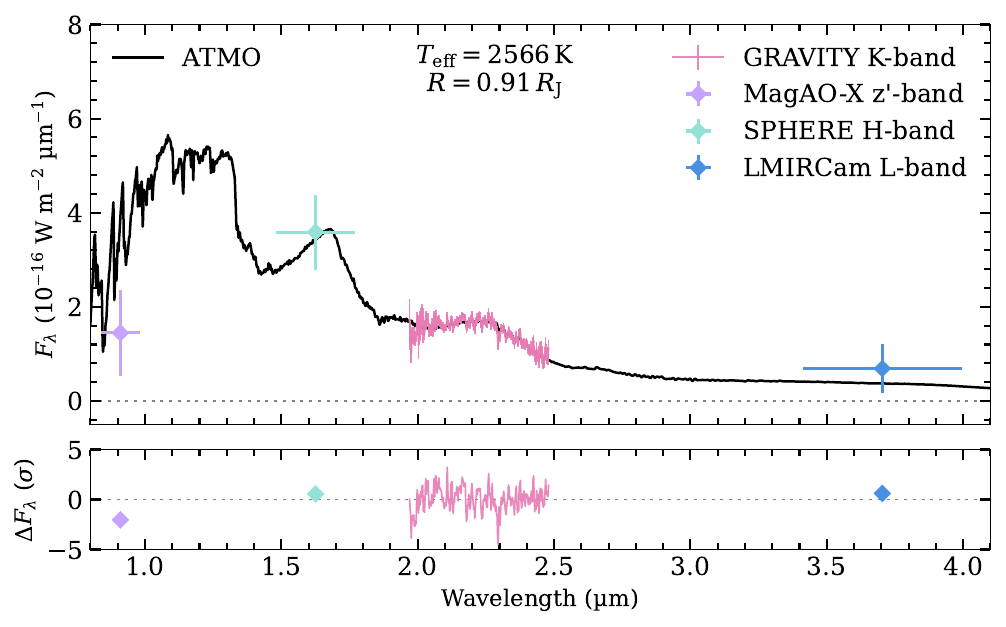}{0.48\textwidth}{(f)}
}

\caption{
Exploration of multiple model grids.
Panels show best-fit spectra from 
(a) Sonora-Diamondback,
(b) ExoRem,
(c) BT-Dusty,
(d) Sonora-Bobcat,
(e) BT-Settl,
and (f) ATMO,
with inferred planet temperatures and radii.
}

\label{spectrums}
\end{figure*}

We do not attempt to constrain these values as the continuum shape is only marginally influenced by log \textit{g} and we do not resolve any individual spectral lines that would be strongly influenced (see Fig.~\ref{logg}).
Because the GRAVITY spectrum has a moderate spectral resolution, the $K$-band spectrum is not sensitive enough to the depth of lines to determine the abundances of individual elements \citep{Rukdee2024} (see Fig.~\ref{metallicity}). For this reason we fix the metallicity at 0.0, assuming a solar value.
The resulting posterior distributions for out best fit \texttt{Drift-Phoenix} model are shown in Fig.~\ref{cornerplot_drift}.
Additional analysis determined that the atmospheric constraints are dominated by the $K$-band spectrum.
Starting from the $K$-band spectrum we ran our sampler to extract the planet parameters and adding $L$-, $H$-, and $z'$ in that order.
We find this produces only a very minimal change in temperature, on the order of tens of kelvin, letting us include all photometric points in the final fits.
We show in Fig.~\ref{spectrums}, a comparison of our test models. 
As the fits with the smallest residuals appear to be \texttt{Drift-Phoenix} and \texttt{ExoRem}, this may imply a cloudy atmosphere. 
Although we do note that both \texttt{BT-Settl} and \texttt{Sonora-Diamondback} also employ cloud physics but are not ideal fits to the data.

\FloatBarrier 


\FloatBarrier 


\FloatBarrier 


\FloatBarrier 


\FloatBarrier 
\clearpage

\end{document}